\documentclass[aps,pra,twocolumn,showpacs]{revtex4-1}
\usepackage{graphicx}
\usepackage{amsmath,amsfonts,amssymb}
\usepackage{CJK}
\usepackage{color, soul}
\setulcolor{red}
\begin{document}
\begin{CJK*}{GBK}{song}

\title{Understanding tunneling ionization of atoms in laser fields with the
essence of multiphoton absorption}
\thanks{Supported by National Natural Science Foundation of China
(No. 11725417,11575027), NSAF (Grant No. U1730449), and Science Challenge Project
(Grant No. TZ2018005).}

\author{Long Xu}
\affiliation{Graduate School of China Academy of Engineering Physics, No. 10 Xibeiwang
East Road, Haidian District, Beijing, 100193, China}

\author{Li-Bin Fu}
\email{Corresponding author. Email: lbfu@gscaep.ac.cn}
\affiliation{Graduate School of China Academy of Engineering Physics, No. 10 Xibeiwang
East Road, Haidian District, Beijing, 100193, China}

\begin{abstract}
The elaborate energy and momentum spectra of ionized electrons from atoms in laser fields
suggest that the ionization dynamics described by tunneling theory should be modified.
Although many efforts have been done within semiclassical models,
there are few discussions describing multiphoton absorption process with quantum framework.
In this letter, by comparing the results obtained with
the time-dependent Schr\"{o}dinger equation (TDSE) and Keldysh-Faisal-Reiss (KFR) theory,
we have studied the nonperturbative effects of ionization dynamics beyond KFR theory.
The difference in momentum spectra between multiphoton and tunneling regimes
is understood in a unified picture with virtual multiphoton absorption processes.
For the multiphoton regime, the momentum spectra can
be obtained by coherent interference of each periodic contribution. However,
the interference of multiphoton absorption peaks will result in the complex
structure of virtual multiphoton bands in the tunneling regime.
It is shown that the virtual spectra will be almost continuous in the tunneling regime
instead of the discrete levels in the multiphoton regime.
Finally, with a model combining TDSE and KFR theory, we have tried to understand the different
effects of virtual multiphoton processes on ionization dynamics.

\end{abstract}
\maketitle

The behavior of electron in a strong laser field has attracted considerable interest,
especially after the observation of above-threshold ionization (ATI) \cite{Agostini1979},
where the electron can absorb more photons than that required to overcome the ionization potential.
Since then, a variety of experiments have been performed in order to reveal the
underlying mechanisms of various phenomena appeared in the strong field,
such as the ATI \cite{Freeman1987,Schafer1993,Milosevic2010,Gong2015},
tunneling ionization \cite{Augst1989,Urbain2004,Wu2012,Camus2017}, and
high-order harmonic generation (HHG) \cite{Ferray1987,McPherson1987,Winterfeldt2008}.

The interaction of atoms with strong laser fields can be studied numerically with
TDSE \cite{Parker1996,Tong1997,Lein2000,Zielinski2016}, classical
 \cite{Paulus1994,Panfili2002} or semiclassical approaches \cite{Corkum1989,
Corkum1993,Chen2000}.
By using semiclassical methods, we can understand the profound
physical processes and come to universal results. Here, most of
the semiclassical methods are based on the KFR theory
\cite{Keldysh1965,Faisal1973,Reiss1980,Popruzhenko2014},
which can result in qualitative agreement with the
results of exact TDSE or experiments, e.g. the energy spectra of ATI
\cite{Milosevic2006,Wickenhauser2006}.
KFR theory ignores the dynamics of the bound states and the coulomb effect on the continuum states.
In the past, many efforts have been made in the KFR theory's amendments, which mainly considers
the Coulomb interaction, such as Coulomb-Volkov approximation \cite{Reiss1994,Arbo2008,Faisal2016}
and the rescattering models of Strong Field Approximation
\cite{Becker1994}. The effects of the excited states
 \cite{Smirnova2006,Serebryannikov2016} and the depletion effect of the
ground state \cite{Milosevic2010,Lewenstein1994} have also been evaluated
within the KFR theory.

In the pioneering paper of KFR theory, Keldysh first proposed the so-called Keldysh
parameter \cite{Keldysh1965} defined as $\gamma \equiv \sqrt{I_P/2U_P}$,
where $I_P$ is the ionization potential and $U_P=E_{0}^{2}/4\omega ^{2}$
is the ponderomotive energy. Here $E_{0}$ and $\omega $ are the amplitude and
frequency of laser pulse, respectively. The asymptotic
behavior of the ionization rate at $\gamma \rightarrow 0$ tends to the case of
electron tunneling in a static field \cite{Landau1977}.
Hence when $\gamma \ll 1$, the ionization process is known as the tunneling
regime, whereas it is regarded as the multiphoton regime for $\gamma \gg 1$ \cite%
{Gontier1980,Fabre1982}, although both of their processes are related to
multiphoton absorption.
It's widely accepted that KFR theory only works in the tunneling regime.
Besides, both the semiclassical methods \cite{Corkum1989,Corkum1993,Chen2000} and
the rescattering models of Strong Field Approximation \cite{Becker1994} work
very well in the tunneling regime where the rescattering effect dominates the dynamics.

Most recently, considerable amount of attention are paid to ionization dynamics in the
crossover regime ($\gamma \sim 1$), where nonadiabatic effects
\cite{Yudin2001,Boge2013,Klaiber2015} and modifications of the semiclassical
methods \cite{Shvetsov-Shilovski2016,Li2014} have been discussed.
These discussions are triggered by the fact that the momentum spectra
observed in the experiments indicate that the tunneling ionization theory
should be modified. Many efforts combining multiphoton and tunneling
processes have also been made to understand the ionization process in
the tunneling and crossover regimes \cite{Serebryannikov2016,Klaiber2016}.
In these regimes, the rescattering effect dominates the dynamics and
covers the other nonperturbative effects. Hence, to investigate the
modifications of ionization dynamics beyond KFR theory, we need a unified
model that excludes rescattering process to bridge the multiphoton and tunneling regimes.

In this letter, by solving TDSE and comparing the results with KFR
theory, we find the virtual multiphoton absorption processes play an
important role in understanding multiphoton and tunneling ionization
dynamics in a unified picture. Then, with a model that combines TDSE and KFR
theory, we investigate the nonperturbative effects of ionization dynamics other than
the rescattering effect. Based on such a model, we illustrate
different nonperturbative effects by filtering the contribution of the
virtual multiphoton process and understand them essentially from the
multiphoton absorption.

Without loss of generality, we consider a one dimensional model
(in atomic units):
\begin{equation}\label{TDSE}
i\frac{\partial}{\partial t} \Phi (x,t)=\left[ -\frac{1}{2}\frac{%
\partial ^{2}}{\partial x^{2}}-\frac{1}{\sqrt{x^{2}+a^2}} +x E(t)\right] \Phi (x,t),
\end{equation}
where the second and third terms in the right hand represent the soft-core electron-nucleus interaction and the atom-field interaction,
respectively. The soft-core parameter is chosen as $a^2=0.484$ in order
to reproduce singly charged helium atom. The pulse is designed as $E(t)=-\partial A(t)/\partial t$ with the vector potential $
A(t)=E_0/\omega \sin^2 (\pi t/\tau)\cos(\omega t)$,
with $\tau$ as the pulse duration.
In the simulation, a total of 16 optical cycles are used unless stated otherwise.
In the numerical calculations, we employ the split-operator method \cite{Feit1982} to solve TDSE and imaginary-time propagation to compute the singlet ground state as the initial wave function.

Furthermore, we can expand the unknown wave function in terms of the basis combining the bound wave function $\varphi_n (x,t)$ and Volkov function $\psi_p (x,t)=(2\pi)^{-1/2} \exp\{i[p+A(t)]x-i
\int_{0}^{t}[p+A(t^{\prime })]^{2}/2 d t^{\prime }\}$ as
\begin{eqnarray}\label{expansion}
\Phi(x,t)
= \sum_p a_p(t)\psi_p(x,t) + \sum_n b_n(t)\varphi_n(x,t).
\end{eqnarray}
Then, substituting the expansion $\eqref{expansion}$ into the Schr\"{o}dinger equation
$\eqref{TDSE}$ and making assumptions: (a) Coulomb interaction is neglected
for continuum states; (b) the population of all the excited states is ignored, we have
\begin{eqnarray}
i\frac{\partial}{\partial t}a_p(t)
=  b_{g}(t) \langle \psi_p(x,t)|x E(t)|\varphi_{g}(x,t)\rangle,
\label{modelA}\\
i\frac{\partial}{\partial t}b_{g}(t)
= -\sum\limits_p a_p(t) \langle \varphi_{g}(x,t)|\frac{1}{\sqrt{x^{2}+a^2}}|\psi_p(x,t)\rangle.
\label{modelB}
\end{eqnarray}

Moreover, after making the assumption (c): the dynamics of ground state is neglected,
we will obtain the final ionization amplitude, which retains the original
choice of Keldysh \cite{Keldysh1965},
\begin{eqnarray}\label{KFR}
i \frac{\partial}{\partial t}a_p(t) = \langle\psi_p(x,t)|x E(t)|\varphi_g(x,t)\rangle.
\end{eqnarray}

\begin{figure}[b]
\centering
\includegraphics[width=\linewidth]{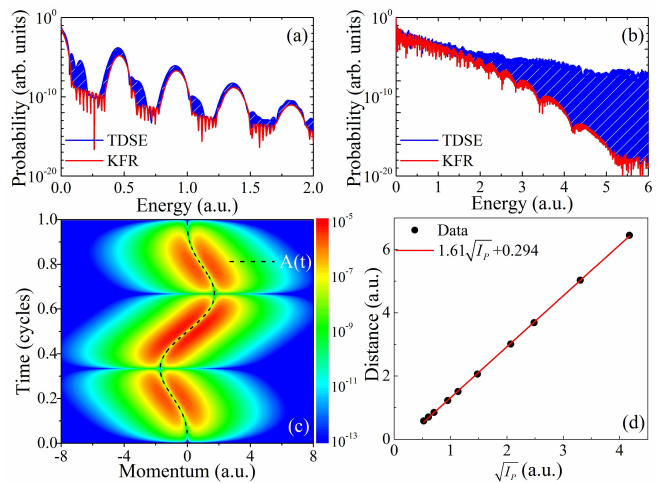}
\caption{Normalized energy distributions calculated by TDSE (blue) and KFR theory (red)
for the wavelength are (a) 100 nm (Keldysh parameter $\gamma = 8$),
(b) 1600 nm (Keldysh parameter $\gamma = 0.5$), respectively, with the laser
intensity is 2$\times 10^{14}\mathrm{W/cm^{2}}$, where blue shading areas represent the
difference between TDSE and KFR theory calculations.
(c) $|\langle \protect\psi _{p}(x,t)|x E(t)|\protect\varphi _{g}(x,t)\rangle |^{2}$
as a function of momentum and time for the case of (b), where the
dashed line represents the vector potential A(t).
(d) The distance between two peaks of
$|\langle \protect\psi _{p}(x,t)|x|\protect\varphi _{g}(x,t)\rangle |^{2}$ is
plotted as a function of $\sqrt{I_P}$, i.e., the 0.5th power of ionization potential.}
\label{figure1}
\end{figure}

Keldysh arrived at the ionization rate by integrating the momentum
distribution (see Eq. $\eqref{KFR}$). After using Fourier series
expanding the expression and using the saddle-point method calculating the
integration, the probability of direct ionization from the ground state was
obtained, where factor $\gamma $ was introduced \cite{Keldysh1965}. KFR theory
can describe the ionization rate well, but it fails to predict the
structures of energy distribution, that is, the positions of energy peaks are
completely different from the results of TDSE when $\gamma $ is
small (see Figs. \ref{figure1}(a) and \ref{figure1}(b)).
Rescattering events mainly affect the yield of high-energy
photoelectrons (energy is greater than 2$U_P$).
Besides the rescattering effect, there
are still other nonperturbative effects which lead KFR theory to deviate
from the actual results for small $\gamma $.

\begin{figure}[t]
\centering
\includegraphics[width=7.5cm,height=5cm]{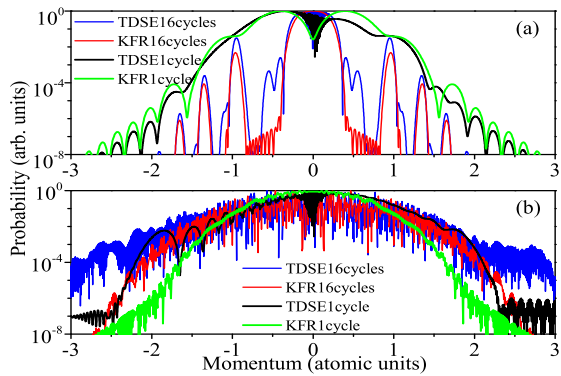}
\caption{Normalized momentum distributions for the cases shown in Figs. \ref{figure1}(a) and \ref{figure1}(b).
Blue and black lines represent the results of TDSE for 16 optical cycles (blue) and one cycle (black),
while red and green lines denote the results of KFR theory for 16 cycles (red) and one cycle (green), respectively.}
\label{figure2}
\end{figure}

To gain the physical insights over the Keldysh parameter, we will
study the exact ionization dynamics with different $\gamma$.
Firstly, we calculate the integrand of Eq. $\eqref{KFR}$ and
plot the square of its module, namely,
$|\langle \psi _{p}(x,t)|x E(t)|\varphi _{g}(x,t)\rangle |^{2}$,
in Fig. \ref {figure1}(c).
At a given moment, $|\langle \psi _{p}(x,t)|x E(t)|\varphi _{g}(x,t)\rangle |^{2}$
as a function of momentum shows a bimodal structure,
which does not change with the time and its center position moves with $A(t)$ between $-A_0$ and $A_0$,
that is, the variation of offset caused by laser field is $2A_0=4\sqrt{U_{P}}$.
Additionally, we scan the ionization potential by choosing different
soft-core parameters and plot the distance between two peaks of the bimodal
structure as a function of $\sqrt{I_P}$ in Fig. \ref{figure1}(d),
where the distance equals $1.61\sqrt{I_P}+0.294$.
Fig. \ref{figure2} shows the momentum distributions of one and $16$ optical cycles.
For the energy spectra of one cycle, both the results of TDSE and KFR theory
show similar results with bimodal structures at $\gamma =8 $ and
arched structures at $\gamma =0.5 $.
The momentum spectrum of one cycle is a bimodal structure for large $\gamma$.
And when the offset variation $4\sqrt{U_{P}}$ of the center position of
$|\langle \psi _{p}(x,t)|x E(t)|\varphi _{g}(x,t)\rangle |^{2}$
is greater than the distance $1.61\sqrt{I_P}+0.294$ between the two peaks of $|\langle \psi _{p}(x,t)|x E(t)|\varphi _{g}(x,t)\rangle |^{2}$,
namely, $\gamma <1.76$, the momentum
distribution of one cycle will be merged into an arched structure.
Clearly, the two extreme cases are consistent with the usual classification using the Keldysh parameter: the tunneling regime at $\gamma \ll 1$ and the multiphoton regime at $\gamma \gg 1$.

Figure \ref{figure2}(a) shows that the results of 16 cycles
calculated by TDSE and KFR theory in the multiphoton regime are similar
(also can be seen in Fig. \ref{figure1}(a)), and their envelopes can be
described by the results of one cycle, and these are consistent in TDSE and KFR.
In the tunneling regime, the momentum distribution of KFR theory runs
away from TDSE. Comparing with the clear ATI peaks in the
multiphoton regime, the energy spectrum in the tunneling regime is more
complicated although both of them are multiphoton absorption.
The blue shading areas in Fig. \ref{figure1} show the difference
between the results of TDSE and KFR theory, and the difference results from
the dynamics of all the bound states and the coulomb effect on the continuum states.
The rescattering takes over the behavior of high-energy spectrum for $\gamma \ll 1$,
which has been studied essentially \cite{Becker1994}.
In addition to the rescattering effect, however, there should
be other nonperturbative effects, which are not included
in the KFR theory and semiclassical models \cite{Corkum1989, Corkum1993, Chen2000},
also contribute to the blue shading areas.

\begin{figure}[t]
\centering
\includegraphics[width=\linewidth]{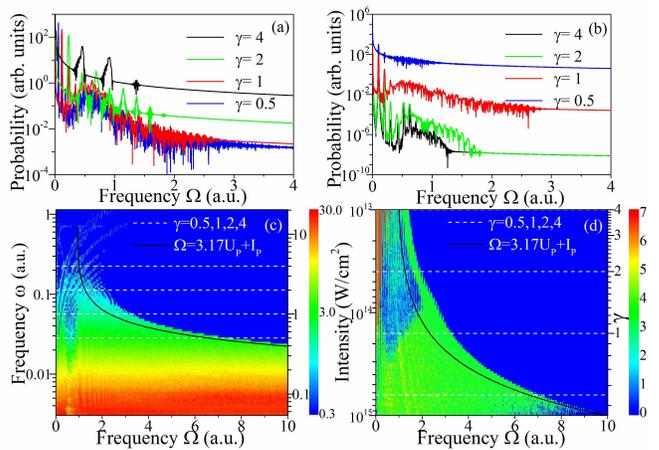}
\caption{Frequency spectrum analysis of the ground-state population for
different Keldysh parameters. (a) The laser intensity is fixed as 2$\times
10^{14}\mathrm{W/cm^{2}}$, and the number of oscillations is plotted in (c). (b) The
laser frequency is kept as 0.05 a.u., and the number of oscillations is plotted
in (d). The solid lines in (c) and (d) guide the structures of the maximum of
frequency $\Omega $ as a function of Keldysh parameter.}
\label{figure3}
\end{figure}

\begin{figure}[t]
\centering
\includegraphics[width=\linewidth]{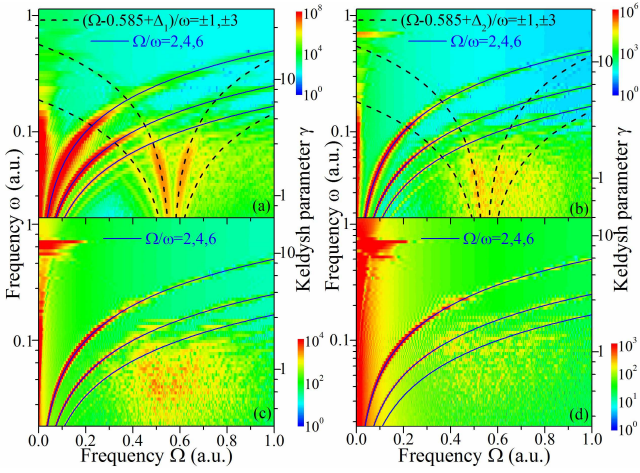}
\caption{Fourier spectra of ground-state population with the laser
frequency is scanned from $0.02$ to $1.1$ a.u. for different intensities:
(a) 0.5, (b) 1, (c) 2, (d) 4 $\times 10^{14}\mathrm{W/cm^{2}}$, respectively.
Normalized by the corresponding final probability $F(E_{0},\omega,\Omega=\infty )$.
Blue solid lines represent the frequency of Fourier spectra that satisfies $\Omega/\omega=2, 4, 6$,
while black dashed lines guide the structures of virtual photons correlated with
the first excited state in (a) and (b).}
\label{figure4}
\end{figure}

Besides the rescattering, the evolution of the ground-state has also
been ignored in the KFR theory and it should contain important
information about dynamics.
Here, we apply Fourier Transform
$F(E_{0},\omega ,\Omega )=(2\pi )^{-1} \int c_{g}(E_{0},\omega ,t)e^{-i\Omega t}dt$
to extract the information
carried by the ground-state population
$c_{g}(E_{0},\omega ,t)= |\langle \varphi_{g}(x,t)|\Phi(x,t)\rangle|^2$,
where $\Omega$ is the frequency of Fourier spectra.
The results of frequency spectrum analysis are plotted in Fig. \ref{figure3}.
Figs. \ref{figure3}(a) and \ref{figure3}(b) show significantly enhanced probability in the region of $\Omega$ between
0.5 and 1 a.u. and this is related to the excited states and details will be shown
in Fig. \ref{figure4}.
Additionally, there are always even laser frequencies in the spectra (also can be
seen in Fig. \ref{figure4}), which are the behaviors of virtual multiphoton processes.
According to Floquet's theory \cite{Shirley1965}, there are Floquet states by
emitting or absorbing integer photons from bound states in the presence of
periodic laser field and the wave function changes parity after absorbing or
emitting a photon as seen from dipole matrix element.
Judging from the form of the ground-state population,
we know that only the states whose parity is the same as the ground state will show
in the ground-state population.
With decreasing $\gamma $, the peaks of virtual multiphoton absorption are
broadening and interfering with each other, resulting in the Fourier spectra
become incoherent and chaotic.
These features can be found clearly in Figs. \ref{figure3}(c) and \ref{figure3}(d) where
we count the numbers of peaks of Fourier spectra and plot the numbers as a
function of $\gamma$.
Clearly, the range of oscillation can be depicted as $\Omega =3.17U_{P}+I_{P}$.
which agrees with the cutoff of HHG \cite{Schafer1993, Lewenstein1994},
shows that the maximum energy of the electron
returns to the ground state is $3.17U_{P}$ as well.

The above Fourier spectra show that the incoherent and chaotic structures
resulting from the interference between virtual multiphoton absorption peaks.
Additionally, the energy of interaction is $d\cdot E(t)$, where $d$ is the dipole moment,
and the separation of adjacent peaks is photon frequency $\omega$.
At the region $d\cdot E_0 \gg \omega$ ($ \gamma \ll 1$),
the peaks interfere with each other.
In other words, if laser intensity is kept constant and $\omega$ becomes smaller,
the increasing Floquet states make the separation to be smaller.
If $\omega$ is kept constant and the laser intensity becomes higher, namely,
the energy of interaction is larger, the more states will be coupled.
Consequently, owing to the more states are involved in the dynamics, the interference
between virtual multiphoton processes will be incoherent, leading to the chaos
of spectrum.

We know that the real chaotic structure is caused by incommensurate
frequencies are involved. To show this clearly, we discuss the features of
the region near the first excited state, $\Omega =0.585a.u.$  \cite{Feit1982}. We
obtain the ground-state population by scanning the laser frequency for four
different laser intensities and plot the Fourier spectra of the population in
Fig. \ref{figure4}, which are normalized by $F(E_{0},\omega,\Omega=\infty )$.
In Fig. \ref{figure4}, we use $\Omega/\omega =2, 4, 6$ and
$(\Omega-0.585)/\omega =\pm 1, \pm 3$ to mark
 the two groups of data, where $2, 4, 6$ and $\pm 1, \pm 3$ are due to the parity.
Fig. \ref{figure3} shows that even laser frequencies come from the Floquet states related
to the ground state and the features of the region near $%
\Omega =0.585a.u.$ result from the contribution of the Floquet states
correlated with the first excited state.
We know that one of the AC Stark shift is proportional to $\alpha E_0^2$ \cite{Delone1999}, where
$\alpha$ represents the static polarizability of an atom, so we choose $\Delta_1= 0.02 a.u.$
and $\Delta_2= 0.04 a.u.$ in order to match the structures.
Additionally, with increasing laser
intensity, these features become more complex and chaotic for quite a strong
field, which are attributed to the coupling of more incommensurate
frequencies.

\begin{figure}[!htb]
\centering
\includegraphics[width=\linewidth]{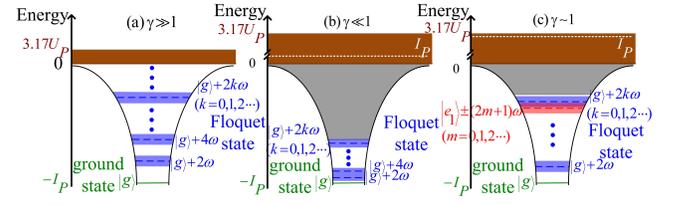}
\caption{Partial diagrammatic sketch of transition between various energy
levels for different Keldysh parameters. Green, gray and brown lines
represent the ground state, intermediate virtual states, and continuum states,
while blue and red lines represent Floquet states correlated with the ground and
first excited state, respectively. Red lines can also represent the effect
of highly excited states. The shadow represents the effect of the spectral
bandwidth and only the state whose parity is the same as the ground state is
plotted in the figure.}
\label{figure5}
\end{figure}

From the above discussions, we combine multiphoton and tunneling ionization
behaviors by coming up with a unified picture with virtual multiphoton
absorption, as shown in Fig. \ref{figure5}.
Fig. \ref{figure5} presents the diagrammatic sketch of transition between
various energy levels for different Keldysh parameters.
When $\gamma $ is large, even laser
frequencies are always present in the frequency spectra (see Figs. \ref{figure3}
and \ref{figure4}), indicating that electron can jump between the
ground state and Floquet states related by virtual multiphoton absorption.
Owing to the spectral line width of the laser pulse, there are
discrete bands of energy under the threshold, instead of energy levels (see
Fig. \ref{figure5}(a)). On the contrary, when $\gamma $ is small, the frequency
spectra will be chaotic, as shown in Figs. \ref{figure3}(c) and \ref{figure3}(d),
illustrating that the whole states will be connected together into a single
stretch (see Fig. \ref{figure5}(b)). In Fig. \ref{figure5}(b), the gray area
represents the intermediate virtual states formed by coupling the continuum
states with the laser field, and the lowest energy is determined by the
laser strength $E_{0}$. Under the level of lowest energy, the Floquet states
associated with ground state link up into a single stretch as well, mainly
due to the interference of virtual multiphoton absorption
peaks. Finally, the behavior of electron appears as tunneling from the ground state.
For the crossover regime ($\gamma \sim 1$), as presented in Fig. \ref{figure5}(c),
there are still discrete energy bands under the lowest energy and
correspondingly, the behavior of electron is shown as tunneling from a virtual
excited state or real excited state in the semiclassical image.

\begin{figure}[t]
\centering
\includegraphics[width=6.5cm,height=4.5cm]{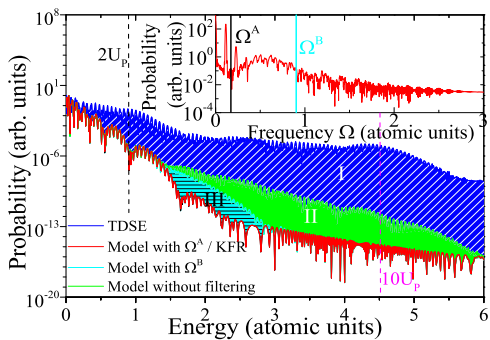}
\caption{Energy distributions calculated by TDSE, KFR theory and the
model with different filtering frequencies.
The laser frequency and intensity are 800 nm, 2$\times 10^{14}\mathrm{W/cm^{2}}$,
respectively.
The results of the KFR theory and the model with $\Omega ^{A}=3\protect\omega$ are completely coincident,
while cyan (Model with $\Omega ^{B}=I_P$) and green (Model without filtering)
lines are the same as the red line for energy lower than 2$U_P$.
The frequency spectrum is plotted in the inset.}
\label{figure6}
\end{figure}

Furthermore, it is meaningful to investigate the performance of different
virtual band structures on the energy spectrum.
Substituting the ground-state population calculated by
TDSE into Eq. \eqref{modelA}, we have
\begin{equation}
i \frac{\partial}{\partial t}a_{p}(t)=\sqrt{c_{g}(t)}\langle \psi
_{p}(x,t)|x E(t)|\varphi _{g}(x,t)\rangle.
\label{model2}
\end{equation}
To investigate the contribution of different virtual bands, we
apply the Fourier transform to $c_{g}(t)$ and obtain the new
population $c_{g}^{\prime }(t)$ by performing the inverse Fourier
transform on the filtered spectrum. In
other words, the new population with filtering can be obtained by $%
c_{g}^{\prime }(t)=\left\vert (2\pi )^{-1}\int\int  c_{g}(t)e^{-i\Omega t}dtf^{K}(\Omega
)e^{i\Omega t}d\Omega \right\vert $, where $f^{K}(\Omega )=
1$ (if $\ |\Omega| \leq \Omega ^{K}$) or $0$ (otherwise)
 is the filtering function with the filtering frequency $\Omega ^{K}$.
We use $f^K(\Omega)$ with different smoothing decay and the results are
qualitatively consistent.
The results of the model with different filtering frequencies are plotted in
Fig. \ref{figure6}, in which we filter the frequencies of the population $c_{g}(t)$ for three cases.

Apparently, the result of TDSE is fairly different from others because it
contains all the information about dynamics and the rescattering effect
covers the other effects.
Besides, the result of the model with $\Omega ^{K}=3\omega$
is the same as the KFR theory, while the other filtering results are similar to
KFR theory only in the region of energy lower than 2$U_P$.
Comparing with the result of KFR theory, the filtering
results are several orders of magnitude higher for energy greater than 2$U_P$.
Here the energy distribution with $\Omega ^{K}=I_P$ only affects a small region
and energy of region improved by the result without filtering is even up to 10$U_P$.
As shown in Fig. \ref{figure6},
the energy distribution is divided into three parts, where region I represents
the behavior of rescattering, while regions II and III represent the nonperturbative
effects of chaotic virtual multiphoton bands and excited
states, respectively.

In summary, we find that, even with the rescattering process ignored, the momentum spectra for
both the multiphoton and tunneling regimes are still quite
different. Such difference is caused by the nonperturbative effects of
ionization dynamics beyond KFR theory, such as the dynamics of the ground state.
In the multiphoton regime, the
momentum spectra of many cycles can be obtained by coherent interference of
contributions from each cycle. However, during the interference of
intercycle, the breadths of multiphoton absorption peak will result in the
complex structure in the tunneling regime.
Moreover, the virtual spectra are almost continuous in the tunneling regime
instead of the discrete bands in the multiphoton regime.
The virtual multiphoton
absorption processes play an important role in understanding multiphoton and
tunneling ionization dynamics. Finally, we put forward a model to understand
the different effects of virtual multiphoton processes on the ionization
dynamics.


 \end{CJK*}

\begin{thebibliography}{99}
\bibitem{Agostini1979} \label{Agostini1979}
        Agostini P et al 1979 {\it Phys. Rev. Lett.} {\bf 42} 1127

\bibitem{Freeman1987} \label{Freeman1987}
        Freeman R R et al 1987 {\it Phys. Rev. Lett.} {\bf 59} 1092

\bibitem{Schafer1993} \label{Schafer1993}
        Schafer K J et al 1993 {\it Phys. Rev. Lett.} {\bf 70} 1599

\bibitem{Milosevic2010} \label{Milosevic2010}
        Milo\v{s}evi\'{c} D B et al 2010 {\it J. Phys. B} {\bf 43} 015401

\bibitem{Gong2015} \label{Gong2015}
        Gong X et al 2015 {\it Phys. Rev. Lett.} {\bf 114} 163001

\bibitem{Augst1989} \label{Augst1989}
        Augst S et al 1989 {\it Phys. Rev. Lett.} {\bf 63} 2212

\bibitem{Urbain2004} \label{Urbain2004}
        Urbain X et al 2004 {\it Phys. Rev. Lett.} {\bf 92} 163004

\bibitem{Wu2012} \label{Wu2012}
        Wu J et al 2012 {\it Phys. Rev. Lett.} {\bf 108} 183001

\bibitem{Camus2017} \label{Camus2017}
        Camus N et al 2017 {\it Phys. Rev. Lett.} {\bf 119} 023201

\bibitem{Ferray1987} \label{Ferray1987}
        Ferray M et al 1987 {\it J. Phys. B} {\bf 21} L31

\bibitem{McPherson1987} \label{McPherson1987}
        McPherson A et al 1987 {\it J. Opt. Soc. Am. B} {\bf 4} 595

\bibitem{Winterfeldt2008} \label{Winterfeldt2008}
        Winterfeldt C, Spielmann C and Gerber G 2008 {\it Rev. Mod. Phys.} {\bf 80} 117

\bibitem{Parker1996} \label{Parker1996}
        Parker J, Taylor K T and Clark C W 1996 {\it J. Phys. B} {\bf 29} L33

\bibitem{Tong1997} \label{Tong1997}
        Tong X-M and Chu S-I 1997 {\it Chem. Phys.} {\bf 217} 119

\bibitem{Lein2000} \label{Lein2000}
        Lein M, Gross E K U and Engel V 2000 {\it Phys. Rev. Lett.} {\bf 85} 4707

\bibitem{Zielinski2016} \label{Zielinski2016}
        Zielinski A, Majety V P and Scrinzi A 2016 {\it Phys. Rev. A} {\bf 93} 023406

\bibitem{Paulus1994} \label{Paulus1994}
        Paulus G G et al 1994 {\it J. Phys. B} {\bf 27} L703

\bibitem{Panfili2002} \label{Panfili2002}
        Panfili R, Haan S L and Eberly J H 2002 {\it Phys. Rev. Lett.} {\bf 89} 113001

\bibitem{Corkum1989} \label{Corkum1989}
        Corkum P B, Burnett N H and Brunel F 1989 {\it Phys. Rev. Lett.} {\bf 62} 1259

\bibitem{Corkum1993} \label{Corkum1993}
        Corkum P B 1993 {\it Phys. Rev. Lett.} {\bf 71} 1994

\bibitem{Chen2000} \label{Chen2000}
        Chen J et al  2000 {\it Phys. Rev. A} {\bf 63} 011404

\bibitem{Keldysh1965} \label{Keldysh1965}
        Keldysh L V 1965 {\it Soviet Physics - JETP} {\bf 20} 1307

\bibitem{Faisal1973} \label{Faisal1973}
        Faisal F H M 1973 {\it J. Phys. B} {\bf 6} L89

\bibitem{Reiss1980} \label{Reiss1980}
        Reiss H R 1980 {\it Phys. Rev. A} {\bf 22} 1786

\bibitem{Popruzhenko2014} \label{Popruzhenko2014}
        Popruzhenko S V 2014 {\it J. Phys. B} {\bf 47} 204001

\bibitem{Milosevic2006} \label{Milosevic2006}
        Milo\v{s}evi\'{c} D B et al 2006 {\it J. Phys. B} {\bf 39} R203

\bibitem{Wickenhauser2006} \label{Wickenhauser2006}
        Wickenhauser M, Tong X M and Lin C D 2006 {\it Phys. Rev. A} {\bf 73} 011401

\bibitem{Reiss1994} \label{Reiss1994}
        Reiss H R and Krainov V P 1994 {\it Phys. Rev. A} {\bf 50} R910

\bibitem{Arbo2008} \label{Arbo2008}
        Arb\'{o} D G et al 2008 {\it Phys. Rev. A} {\bf 77} 013401

\bibitem{Faisal2016} \label{Faizal2016}
        Faisal F H M 2016 {\it Phys. Rev. A} {\bf 94} 031401

\bibitem{Becker1994} \label{Becker1994}
        Becker W, Lohr A and Kleber M 1994 {\it J. Phys. B} {\bf 27} L325
        \item[]
        Lohr A et al 1997 {\it Phys. Rev. A} {\bf 55} R4003
        \item[]
        Becker W et al 2002 {\it Adv. At. Mol. Opt. Phys.} {\bf 48} 35

\bibitem{Smirnova2006} \label{Smirnova2006}
        Smirnova O, Spanner M and Ivanov M 2006 {\it J. Phys. B} {\bf 39} S307

\bibitem{Serebryannikov2016} \label{Serebryannikov2016}
        Serebryannikov E E and Zheltikov A M 2016 {\it Phys. Rev. Lett.} {\bf 116} 123901

\bibitem{Lewenstein1994} \label{Lewenstein1994}
        Lewenstein M et al 1994 {\it Phys. Rev. A} {\bf 49} 2117

\bibitem{Landau1977} \label{Landau1977}
        Landau L D and Lifshitz E M 1977 {\it Quantum Mechanics} (Oxford: Pergamon)

\bibitem{Gontier1980} \label{Gontier1980}
        Gontier Y and Trahin M 1980 {\it J. Phys. B} {\bf 13} 4383

\bibitem{Fabre1982} \label{F. Fabre1982}
        Fabre F et al 1982 {\it J. Phys. B} {\bf 15} 1353

\bibitem{Yudin2001} \label{Yudin2001}
         Yudin G L and Ivanov M Y 2001 {\it Phys. Rev. A} {\bf 64} 013409

\bibitem{Boge2013} \label{Boge2013}
        Boge R et al 2013
        {\it Phys. Rev. Lett.} {\bf 111} 103003

\bibitem{Klaiber2015} \label{Klaiber2015}
        Klaiber M, Hatsagortsyan K Z and Keitel C H 2015 {\it Phys. Rev. Lett.} {\bf 114} 083001

\bibitem{Shvetsov-Shilovski2016} \label{Shvetsov-Shilovski2016}
        Shvetsov-Shilovski N I et al 2016
        {\it Phys. Rev. A} {\bf 94} 013415

\bibitem{Li2014} \label{Li2014}
        Li M et al 2014 {\it Phys. Rev. Lett.} {\bf 112} 113002
        \item[]
        Geng J W et al 2015 {\it Phys. Rev. Lett.} {\bf 115} 193001

\bibitem{Klaiber2016} \label{Klaiber2016}
        Klaiber M and Briggs J S 2016 {\it Phys. Rev. A} {\bf 94} 053405

\bibitem{Feit1982} \label{Feit1982}
        Feit M D, Fleck J A Jr and Steiger A 1982 {\it J. Comput. Phys.} {\bf 47} 412

\bibitem{Shirley1965} \label{Shirley1965}
        Shirley J H 1965 {\it Phys. Rev.}   {\bf 138} B979

\bibitem{Delone1999} \label{Delone1999}
        Delone N B and Kra V P 1999 {\it Phys.-Usp.} {\bf 42} 669
 \end{thebibliography}
\end{document}